\documentstyle[12pt,aasms4]{article}
\begin{document}
\title{Detection of interstellar H$_2$D$^+$ emission}
\author{Ronald Stark}
\affil{Max-Planck-Institut f\"ur Radioastronomie, 
Auf dem H\"ugel 69, D-53121 Bonn, Germany}
\and
\author{Floris F.S. van der Tak \& Ewine F. van Dishoeck}
\affil
{Sterrewacht Leiden, 
Postbus 9513, NL-2300 RA Leiden, The Netherlands}
\begin{abstract}
We report the detection of the $1_{10}-1_{11}$ ground state transition
of ortho-H$_2$D$^+$ at 372.421 GHz in emission from the young stellar
object NGC 1333 IRAS 4A. Detailed excitation models with a power-law
temperature and density structure yield a beam-averaged H$_2$D$^+$
abundance of $3 \times 10^{-12}$ with an uncertainty of a factor of 
two. The line was not detected toward
W~33A, GL 2591, and NGC 2264 IRS, in the latter source at a level
which is $3-8$ times lower than previous observations.
The H$_2$D$^+$ data provide direct evidence in support of
low-temperature chemical models in which H$_2$D$^+$ is enhanced by the
reaction of H$_3^+$ and HD. The H$_2$D$^+$ enhancement toward NGC 1333
IRAS 4A is also reflected in the high DCO$^+$/HCO$^+$ abundance ratio.
Simultaneous observations of the N$_2$H$^+$ $4-3$ line show that its
abundance is about $50-100$ times lower in NGC 1333 IRAS 4A than in
the other sources, suggesting significant depletion of N$_2$.  The
N$_2$H$^+$ data provide independent lower limits on the H$_3^+$
abundance which are consistent with the abundances derived from
H$_2$D$^+$. The corresponding limits on the H$_3^+$ column density
agree with recent near-infrared absorption measurements
of H$_3^+$ toward W 33A and GL 2591. 
\end{abstract}

\keywords{ISM: abundances --- ISM: molecules --- 
molecular processes --- 
radio lines: ISM --- submillimeter}
\clearpage
\section
{Introduction}

The recent detection of the H$_3^+$ ion in interstellar clouds through
its infrared vibration-rotation lines (Geballe \& Oka 1996; McCall et
al. 1998) is an important confirmation of the gas-phase chemical
networks (Herbst \& Klemperer 1973; Watson 1973). Because of its
symmetry, H$_3^+$ has no allowed rotational transitions contrary to
its deuterated isotopomer H$_2$D$^+$ which has a large permanent
dipole moment (Dalgarno et al. 1973). Thus, H$_2$D$^+$ is important as
a tracer of H$_3^+$ with transitions that can be searched for in
emission. In addition, it is widely believed to play a pivotal role in
the interstellar ion-molecule chemistry at low temperatures where
significant enhancement of deuterated molecules occurs as a result of
fractionation (e.g.
Herbst 1982; Millar et al. 1989). This
process is initiated by the isotope exchange equilibrium reaction ${\rm H}_3^+
+ {\rm HD} \rightleftharpoons {\rm H}_2{\rm D}^+ + {\rm H}_2 \ (1)$
which is shifted in the forward direction at low temperatures (Smith
et al. 1982; Herbst 1982). The formation of H$_2$D$^+$ is followed by
deuterium transfer reactions with e.g. CO to form DCO$^+$, and the
H$_2$D$^+$ enhancement is reflected in the observed large abundance
ratios of e.g. DCO$^+$/HCO$^+$, NH$_2$D/NH$_3$, and DCN/HCN in cold clouds
(e.g. Wootten 1987; Butner
et al.  1995; Williams et al. 1998). 

Over the last 20 years numerous attempts have been made to detect the
$1_{10}-1_{11}$ ortho-H$_2$D$^+$ and $1_{01}-0_{00}$ para-H$_2$D$^+$
ground state lines at 372 and 1370 GHz, respectively. These searches
have mainly been done with the {\it Kuiper Airborne Observatory} (KAO)
(Phillips et al. 1985; Pagani et al. 1992a; Boreiko \& Betz 1993), and
a possible absorption feature at 1370 GHz has been reported by Boreiko
\& Betz (1993) toward Orion. Observations from the ground are very
difficult since the 372 GHz line is at the edge of a strong
atmospheric water absorption line, while the atmosphere at 1370 GHz is
almost completely opaque. With the advent of new submillimeter receivers
equipped with sensitive niobium SIS mixers it has become possible to
search for weak ortho-H$_2$D$^+$ lines from high, dry sites such as
Mauna Kea. Indeed, a ground-based search for this line from the {\it
Caltech Submillimeter Observatory} (CSO) by van Dishoeck et al. (1992)
yielded limits which are up to a factor of hundred more sensitive than
those obtained with the KAO. Comparison with chemical models suggested
that only a factor of a few improvement would be needed to detect the
line. With the new facility receiver RxB3 at the {\it James Clerk
Maxwell Telescope} (JCMT)\footnote{
The JCMT is operated by the Joint Astronomy Centre in Hilo, Hawaii on
behalf of the Particle Physics and Astronomy Research Council in the
United Kingdom, the Netherlands Organisation for Scientific Research,
and the National Research Council of Canada.}
such an improvement in sensitivity is now
achievable. Here we report the detection of the H$_2$D$^+$ 372.421 GHz
line toward NGC 1333 IRAS 4A and significant upper limits toward
W~33A, GL 2591, and NGC 2264 IRS. Simultaneous observations of the
N$_2$H$^+$ $4-3$ line at 372.672 GHz toward these young stellar
objects (YSOs) are used to place additional constraints on the
H$_3^+$ abundance.

\section
{Observations}

The observations of the $1_{10}-1_{11}$ ground state transition of
ortho-H$_2$D$^+$ at 372.42134 GHz (Bogey et al. 1984) were done with the
JCMT on August 31, September 15 and 18, 1998 during three nights of
very good submillimeter transparency with a zenith opacity at 225 GHz
below 0.05. The dual-polarization heterodyne receiver RxB3 was used
(Avery 1998). Both mixers were tuned to 372.5469 GHz in the upper sideband.
The big advantage of RxB3 is that it has a
dual-beam interferometer which allows single-sideband (SSB) operation,
enhancing the sensitivity and calibration at 372 GHz considerably. The
digital autocorrelator spectrometer (DAS) was split into four parts of
125 MHz.
This setup
allows observations of both lines in two orthogonal polarizations
simultaneously with a spectral resolution of 376 kHz ($\equiv 0.3$ km
s$^{-1}$ at 372 GHz). Typical SSB system temperatures including
atmospheric losses were about 1200 K. The effective total integration
time was 7 hours on NGC 1333 observed over two nights, 2.7 hours on
W~33A, 4.3 hours on GL 2591, and 4 hours on NGC 2264. The absolute calibration
uncertainty is estimated at 30\%, and the relative calibration between the
H$_2$D$^+$ and N$_2$H$^+$ lines is much better. The JCMT beamsize at
372 GHz is $13''$ FWHM, the main beam efficiency is 57\%. JCMT data on
H$^{13}$CO$^+$ and DCO$^+$ were taken from the literature (see below), except
for DCO$^+$ toward W~33A and GL 2591, for which the $3-2$ transition at
216.113 GHz was observed with receiver RxA3. The beamsize at this frequency is
21$''$ FWHM and the main beam efficiency is 70\%.

The observed H$_2$D$^+$ and N$_2$H$^+$ spectra are presented in
Figure~1.  The source and line parameters are listed in Table 1. The
H$_2$D$^+$ line is clearly detected with $T_A^*= 0.08 \pm 0.03$ K toward
NGC 1333 IRAS 4A, and is seen in spectra of both nights. The velocity
width shows good agreement with the N$_2$H$^+$ line width while the
velocities are offset by about 0.5 km s$^{-1}$. Comparison with the
line survey of Blake et al.\ (1995) shows that such an offset is small
and common for this region. 
No H$_2$D$^+$
emission was detected toward NGC 2264 IRS, W~33A, and GL 2591.
Assuming the same width as the N$_2$H$^+$ line, 2$\sigma$ upper limits
of $T_A^*\le 0.02 - 0.04$ K are obtained. For NGC 2264 this limit is
about a factor of eight below the possible feature of Phillips et
al. (1985), and a factor of three below the limit reached by van
Dishoeck et al. (1992).
Note that the N$_2$H$^+$ emission toward NGC 1333 is much weaker
than that toward the other sources. No other lines were detected in the 125
MHz bands.

\section
{Analysis}

Model calculations were performed to determine the
abundances of H$_2$D$^+$, N$_2$H$^+$, HCO$^+$, and DCO$^+$
using a power-law density structure $n=n_0 (r/R_o)^{-\alpha}$, as
described in van der Tak et al. (1999). In these models the radial
dust temperature profile is calculated from the observed luminosity
and $n_0$ is determined from submillimeter photometry, which probes
the total dust mass. The grain heating and cooling are solved
self-consistently as a function of radius, $r$, using grain properties
from Ossenkopf \& Henning (1994).  The outer radius ($R_o$) is determined 
from high resolution submillimeter line and
continuum maps.  The exponent $\alpha$ is constrained by modeling
the relative strength of emission lines of CS and H$_2$CO at the
central position over a large range of critical densities with a 
Monte Carlo radiative transfer program, assuming $T_{\rm K}=T_{\rm dust}$. 
Data were taken from Blake et al.\ (1995)
(NGC 1333 IRAS 4A), de Boisanger et al.\ (1996) and Schreyer et al.\
(1997) (NGC 2264 IRS), and van der Tak et al.\ (1999 and in preparation) 
(GL 2591 and W 33A). For NGC 1333 IRAS 4A, where CS is heavily
depleted, $\alpha=2$ was taken based on the analysis of the continuum
visibilities in interferometer data by Looney (1998).

Given the calculated temperature and density structure, the 
radiative transfer models were run to determine the abundances,
assuming initially a constant abundance throughout the envelope. Both 
the ortho-H$_2$D$^+$ and para-H$_2$D$^+$ ladders have been considered
since their spin states are coupled through reactive collisions with
H$_2$; thus the para $0_{00}$ level is the true rotational ground
state. A de-excitation rate coefficient of $1.0 \times 10^{-10}$
cm$^3$ s$^{-1}$ has been used for all inter-ladder transitions.
See Herbst (1982) and Pagani et al. (1992b) for a detailed study of
the ortho/para ratio. The lower level of the $1_{10}-1_{11}$
transition lies at 86 K. The excitation energy of the $1_{10}$ level is
18 K relative to the $1_{11}$ level, and the critical density for this 
transition is about $2\times 10^5$ cm$^{-3}$.

The calculated abundances are listed in Table 2.
Toward NGC 1333 we infer a beam-averaged abundance $x($H$_2$D$^+)=
3 \times 10^{-12}$. Upper limits on the abundance toward 
NGC 2264, W~33A, and GL 2591 are $<1\times 10^{-11}$. The N$_2$H$^+$
abundance ranges between $10^{-11}$ toward NGC 1333 and $10^{-9}$
toward W~33A. All derived abundances have an absolute uncertainty 
of a factor of two, due to the uncertainties in the dust opacities 
and CO abundances. The relatively high N$_2$H$^+$ abundance toward NGC 2264
was already found by van Dishoeck et al.  (1992), who noted that
nearly all of the gas phase nitrogen must be in the form of N$_2$ in
this cloud. Since N$_2$H$^+$ is formed mainly by the reaction of
H$_3^+$ and N$_2$, the latter observations provide an independent
lower limit on the H$_3^+$ abundance.  Destruction occurs mainly via
reactions with CO, O, and electrons. Considering CO destruction only,
$n({\rm H}_3^+)\ga {{0.5~n({\rm N}_2{\rm H}^+)~x({\rm CO})}/{x({\rm
N}_2)}}\ (2).$ Assuming 50\% of the nitrogen is in N$_2$,
$x({\rm N}_2)= 5 \times 10^{-5} \delta({\rm N}_2)$, $x({\rm CO})=  2\times 
10^{-4} \delta({\rm CO})$ and equal amounts of depletion $\delta$ for
CO and N$_2$, this yields $x({\rm H}_3^+)\ga 2 x({\rm N}_2{\rm
H}^+)$. These limits are listed in Table 2, and are consistent with the
upper limits derived from the H$_2$D$^+$ observations using a
theoretical ${\rm H}_2{\rm D}^+/{\rm H}_3^+$ ratio, see \S 4.

\section
{H$_2$D$^+$/H$_3^+$ chemistry}

The above analysis assumes constant abundances throughout the YSO
envelopes.  In reality, the H$_2$D$^+$ abundance is a strong function
of temperature and position.  In chemical
equilibrium, the H$_3^+$ abundance can be written as $x($H$_3^+)=\zeta/
\Sigma k_{\rm X}n({\rm X})$ with $n($X$)=n($H$_2)x($X$)$. X refers to
any of O, C, CO, O$_2$, N$_2$, H$_2$O, ... which are the principal
removal agents of H$_3^+$ via the proton transfer reactions $ {\rm
H}_3^+ + {\rm X} \rightarrow {\rm XH}^+ + {\rm H}_2 \ (3)$, where
$k_{\rm X}$ are the rate coefficients (taken from the UMIST database,
see e.g. Millar et al. 1997) and $\zeta$ is the cosmic-ray ionization
rate (taken to be $5\times 10^{-17}$ s$^{-1}$).  A simple chemical
model for the formation and destruction of H$_2$D$^+$ yields
$$ {x({\rm H}_2{\rm D}^+) \over {x({\rm H}_3^+)}} = 
 {{ x({\rm HD})k_f + x({\rm D}) k_D} \over
        {  x(e) k_e + \sum k_X x({\rm X}) + k_r}}, \eqno(4)$$
where $k_f$ and $k_r$ are the forward and backward rate coefficients
of reaction (1),
$k_D$ is the rate coefficient for formation of H$_2$D$^+$ through the reaction
H$_3^+ + {\rm D}$, and $k_e$ is the rate coefficient
of the electron recombination
of H$_2$D$^+$ (see e.g. Caselli et al.\  1998 for a compilation of
values). We assumed $x({\rm HD})= 10 x({\rm D})= 2.8 \times 10^{-5}$ throughout.

The above H$_3^+$ and H$_2$D$^+$ chemical equations were 
included in the power-law
models, and abundances at each position were calculated for the
appropriate temperature and density. We have fixed the expression for 
$k_r$ at $T<20$ K to its value at 20 K, to ensure
that $x($H$_2$D$^+)< x($H$_3^+)$ throughout. For simplicity, only X=CO was
considered and the electron recombination was neglected. 
The CO depletion is inferred from C$^{17}$O observations
as described in van der Tak et al. (1999 and in preparation), and are listed in 
Table~2.
For a homogeneous temperature and density structure, our model agrees well with 
the models of Millar et al. (1989) and Pagani et al. (1992b).

The power-law model results for NGC 1333 IRAS 4A are presented in Figure~2.   
Using these abundances, the H$_2$D$^+$ emission has been calculated, most of which 
originates from gas at $T=25-35$~K. The model intensity agrees within 30\% with
that measured toward NGC 1333 IRAS 4A, and is consistent with the upper limit toward
GL2591. They are a factor of $2-4$ larger than the upper limits toward NGC 2264 IRS 
and W~33A, respectively. Most likely, this small discrepancy results from effective 
removal of H$_3^+$ by other species than CO and/or by an overestimate of the size of 
the cloud. The H$_3^+$ column densities 
computed for W~33A and GL 2591 agree within a factor of $2-3$ with the directly
observed column densities by Geballe \& Oka (1996) and McCall et al.\
(1998). This good agreement between models and observations provides the
strongest support for the basic H$_3^+$ and H$_2$D$^+$ chemical networks
in dense, cold clouds.

The simple chemical networks described above (Eq.\ (2) and (3)) have
also been used to model the N$_2$H$^+$ abundance from the 
N$_2$H$^+$/H$_2$D$^+$ ratio. Assuming equal amounts of depletion for CO 
and N$_2$, agreement within a factor of two between the measured and modeled 
N$_2$H$^+$ line intensities is obtained. 
The best fitting N$_2$ abundances are included in Table 2.

The derived H$_2$D$^+$/H$_3^+$ ratios are also compared with the
observed DCO$^+$/HCO$^+$ ratios in Table 2.  The enhancement in
H$_2$D$^+$/H$_3^+$ is clearly reflected in the DCO$^+$/HCO$^+$ ratio,
as expected since the latter species are directly formed by reactions
of the former with CO. For NGC 1333 IRAS 4A, DCO$^+$/HCO$^+=0.5$
H$_2$D$^+$/H$_3^+$, where the factor of $0.5$ is consistent with the
statistical branching ratio of $1/3$ within the uncertainties.  The
DCO$^+$/HCO$^+$ ratio toward NGC 1333 IRAS 4A is a factor of
$5-50$ higher than that toward the other sources.  This can be explained
by the difference in physical structure. In cold, dense
(pre-)protostellar cores like NGC 1333 IRAS 4A where the CO and N$_2$
depletions are extreme, the H$_2$D$^+$ abundance is enhanced, because
of the low temperature and because the main removal reactions of
H$_3^+$ are suppressed.  The H$_2$D$^+$ abundance will increase even
stronger than that of H$_3^+$ since reaction (1) becomes the main
destruction channel of H$_3^+$.  In the case of NGC 2264, W~33A, and GL
2591, where the temperatures are higher and the depletion of CO and
N$_2$ is less, reaction (3) becomes the main removal path of H$_3^+$.

The difference in physical structure may have its origin in the
different stages of protostellar evolution. In particular, NGC 1333 IRAS 4A 
has been classified as a Class 0 object (Andr\'e \& Montmerle 1994), and is
thus in a very early stage, with a large
spatial separation between the quiescent and shocked regions (Blake et
al.\ 1995).

\section
{Conclusions}

With the current sensitivity of heterodyne receivers it is now
possible to study the ortho-H$_2$D$^+$ $1_{10}-1_{11}$ 372.421 GHz
line profile in emission in the early stages of star formation deep
inside dense molecular clouds. Its importance lies in the fact that it
is a tracer of H$_3^+$ and that it provides information on the
deuterium abundance and temperature history of a cloud and
on the chemical evolution during star formation. Further observations
of the H$_2$D$^+$ and H$_3^+$ lines in a sample of very young Class 0
and Class I YSOs will therefore be very valuable.

Since the ortho-H$_2$D$^+$ ground state is at 86 K, the
$1_{10}-1_{11}$ line traces both the warm and cold regions, although
the H$_2$D$^+$ enhancement will be strongest in the coldest regions.
Observations of the $1_{01}-0_{00}$ para-H$_2$D$^+$ ground state line
at 1.37 THz toward the continuum of embedded YSOs may reveal cold
H$_2$D$^+$ in absorption. The dual channel {\it German REceiver for
Astronomy at THz frequencies} (GREAT) to be flown on the {\it
Stratospheric Observatory For Infrared Astronomy} (SOFIA) would allow
such observations. Combined with 372 GHz observations from the ground,
the total abundance and the relative population of the ortho- and
para-modifications may be determined which provides information on the
formation, destruction, and excitation processes. Simultaneous deep
observations of the HD $J=1-0$ (2.7 THz) and para-H$_2$D$^+$ ground
state lines toward YSOs may yield a direct measure of the (variation
in) H$_3^+$ abundance over the cloud, and thus of the cosmic-ray
ionization rate.

\acknowledgments
It is a pleasure to thank
Lorne Avery for discussions and test observations at 372 GHz, Henry
Matthews for the non-standard setup for the backend, and Leslie Looney
and Lee Mundy for sharing their model results on NGC 1333 prior to
publication.  The observations would not have been possible without
the flexible observing strategy of Remo Tilanus.  Gerd-Jan van Zadelhoff,
Fred Baas, and Jane Greaves did an excellent job in carrying out these
observations in service. This work is supported by NWO grant
614.41.003.

%
%
\newpage
\figcaption{Observed spectra of H$_2$D$^+$ $1_{10}-1_{11}$ at 372.421 GHz and 
N$_2$H$^+$ $4-3$ at 372.672 GHz. The N$_2$H$^+$ spectra have a resolution of 
0.3 km s$^{-1}$. The H$_2$D$^+$ spectra have been smoothed to 0.6 km s$^{-1}$.}
\figcaption{Power-law model abundances of H$_3^+$, H$_2$D$^+$, and
N$_2$H$^+$ as function of density and temperature.}
%
%
\begin{deluxetable}{lccrll}
\scriptsize
\tablecolumns{7}
\tablewidth{0pc}
\tablecaption{Observations\tablenotemark{a}}
\tablehead{
\colhead{Source\tablenotemark{b}}& \colhead{Molecule}& \colhead{Transition}&
  \colhead{$T_A^*$}&  \colhead{$\Delta V$}& \colhead{$V_{\rm LSR}$}\\ 
\colhead{}& \colhead{}& \colhead{}& \colhead{(K)}&  \colhead{(km s$^{-1}$)}& \colhead{(km s$^{-1}$)}\\
} 
\startdata
N1333&   H$_2$D$^+$& $1_{10}-1_{11}$&   $0.08(0.03)$&     $1.2\pm 0.3$& $\phn7.4\pm 0.2$ \nl
                &   N$_2$H$^+$& $4-3$&             $2.57(0.03)$&       $1.35$&  $\phn6.94$    \nl
N2264&   H$_2$D$^+$& $1_{10}-1_{11}$&   $\le 0.02$\tablenotemark{c}&  &  \nl
                &   N$_2$H$^+$& $4-3$&             $4.51(0.03)$&       $2.67$&  $\phn8.01$    \nl
W33A&   H$_2$D$^+$& $1_{10}-1_{11}$&              $\le 0.04$\tablenotemark{c}&  &    \nl
                &   N$_2$H$^+$& $4-3$&             $3.06(0.06)$&       $4.62$&  $37.40$   \nl
GL2591& H$_2$D$^+$& $1_{10}-1_{11}$&            $\le 0.02$\tablenotemark{c}& &  \nl
                &   N$_2$H$^+$& $4-3$          &   $1.41(0.04)$&       $2.89$&  $-5.82$   \nl
\enddata
\tablenotetext{a}{Values in parentheses represent $1\sigma$ statistical
uncertainties. The absolute uncertainty of the intensity is 30\%; 
$\Delta V$ and $V_{\rm LSR}$ are accurate to better than $0.1$ km s$^{-1}$.}
\tablenotetext{b}{Position (B1950): NGC 1333 IRAS4A: $\alpha=03^{\rm h}26^{\rm m}04\fs8$, 
$\delta=+31\arcdeg03\arcmin14\arcsec$; NGC 2264 IRS: $\alpha=06^{\rm h}38^{\rm m}25\fs0$,
$\delta=+09\arcdeg32\arcmin29\arcsec$;
W33A: $\alpha=18^{\rm h}11^{\rm m}44\fs2$, $\delta=-17\arcdeg52\arcmin56\arcsec$; GL 2591:
$\alpha=20^{\rm h}27^{\rm m}35\fs8$; $\delta=+40\arcdeg01\arcmin14\arcsec$}
\tablenotetext{c}{$2\sigma$ limits: NGC 2264 IRS: $\Delta V=2.5$, GL 2591: $\Delta 
V=3.0$, and W33A: $\Delta V=4.5$ km s$^{-1}$.}
\end{deluxetable}
%
%
\begin{deluxetable}{lllrcccccccr}
\scriptsize
\tablecolumns{12}
\tablewidth{0pc}
\tablecaption{Excitation model parameters and deduced abundances\tablenotemark{a}}
\tablehead{
\colhead{Source}& \colhead{$\alpha$}& \colhead{$n_0$}& \colhead{$T$}&
 \multicolumn{8}{c}{Molecular Abundances} \\
\colhead{}& \colhead{}& \colhead{}& \colhead{$R_i, R_o$}&
\colhead{}& \colhead{}& \colhead{}& \colhead{}& \colhead{}& \colhead{}& \colhead{}&
 \colhead{} \\
\cline{5-12} \\ 
\colhead{}& \colhead{}&  \colhead{(cm$^{-3}$)}& \multicolumn{1}{c}{(K)}&  
  \colhead{H$_2$D$^+$}& \colhead{N$_2$H$^+$}&  
  \colhead{H$_3^+$\tablenotemark{b}}& \colhead{H$_3^+$\tablenotemark{c}}& 
  \colhead{${\rm H}_2{\rm D}^+ \over {\rm H}_3^+$}& 
\colhead{${\rm DCO}^+ \over {\rm HCO}^+\tablenotemark{d}$ }& \colhead{CO\tablenotemark{e}}& 
\colhead{N$_2$}\\
} 
\startdata
N1333& 
2&            $1.7(6)$& $318, 13$&  $\phantom{<}3(-12)$& $1(-11)$& $\phantom{<}2(-10)$& $>2(-11)$&
$\phantom{<}2(-2)$&  $1(-2)$& $4(-6)$& $4(-7)$\nl
N2264& 
1.5&            $1.5(4)$& $293, 18$& $<1(-11)$& $6(-10)$& $<3(-\phantom{0}9)$& $>1(-\phantom{0}9)$& 
$<3(-3)$& $3(-3)$& $6(-5)$& $4(-5)$\nl
W33A&
1&            $2.1(4)$& $280, 16$& $<1(-11)$& $1(-\phantom{0}9)$& $<4(-\phantom{0}9)$&  
                                                                                   $>2(-\phantom{0}9)$& 
$<3(-3)$&            $1(-4)$& $5(-5)$& $2(-5)$\nl
GL2591& 
1.25&           $3.5(4)$& $350, 30$& $<1(-11)$& $5(-10)$&  $<3(-\phantom{0}9)$& $>1(-\phantom{0}9)$& 
$<3(-3)$& $4(-4)$& $2(-4)$& $9(-5)$\nl
\enddata
\tablenotetext{a}{From statistical equilibrium calculations using the appropriate
temperature and density structure as function of distance $r$ to the YSO, 
$n(r)=n_o (r/R_o)^{-\alpha}$ where $R_o$ is the
outer radius of the model envelope: NGC 1333 IRAS 4A: $3.1(3)$ AU, NGC 2264 IRS: $4.7(4)$ AU, W33A: $2.4(5)$ AU,  
GL 2591: $3.1(4)$ AU, and $R_i=R_o/300$ is the inner radius. Notation a(b) indicates $a \times 10^b$.
The accuracy of the deduced abundances is a factor of two.}
\tablenotetext{b}{From H$_2$D$^+$ using a theoretical
H$_3^+/$H$_2$D$^+$ ratio at the effective temperature from which most of the
emission arises (see Figure 2).}
\tablenotetext{c}{From N$_2$H$^+$ analysis (see text).}
\tablenotetext{d}{From H$^{13}$CO$^+$ assuming HCO$^+$/H$^{13}$CO$^+=60$.}
\tablenotetext{e}{From C$^{17}$O assuming CO/C$^{17}$O=2500 and using the
appropriate $N$(H$_2$) from submillimeter dust emission in a 13$''$ beam: NGC 1333 IRAS 4A: 3.1(23) cm$^{-2}$, 
NGC 2264 IRS: 1.2(23) cm$^{-2}$, 
W~33A: 5.2(23) cm$^{-2}$, GL~2591: 1.3(23) cm$^{-2}$.}
\end{deluxetable}
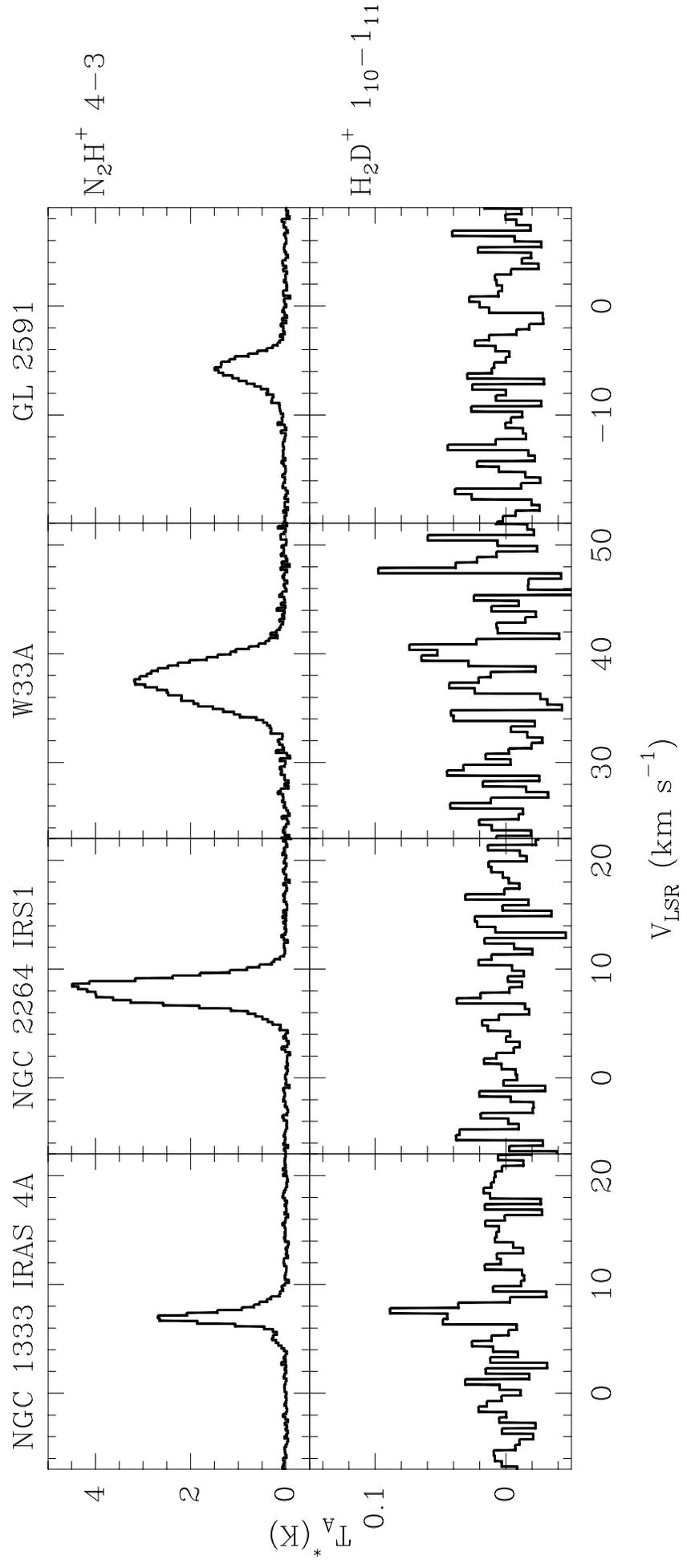
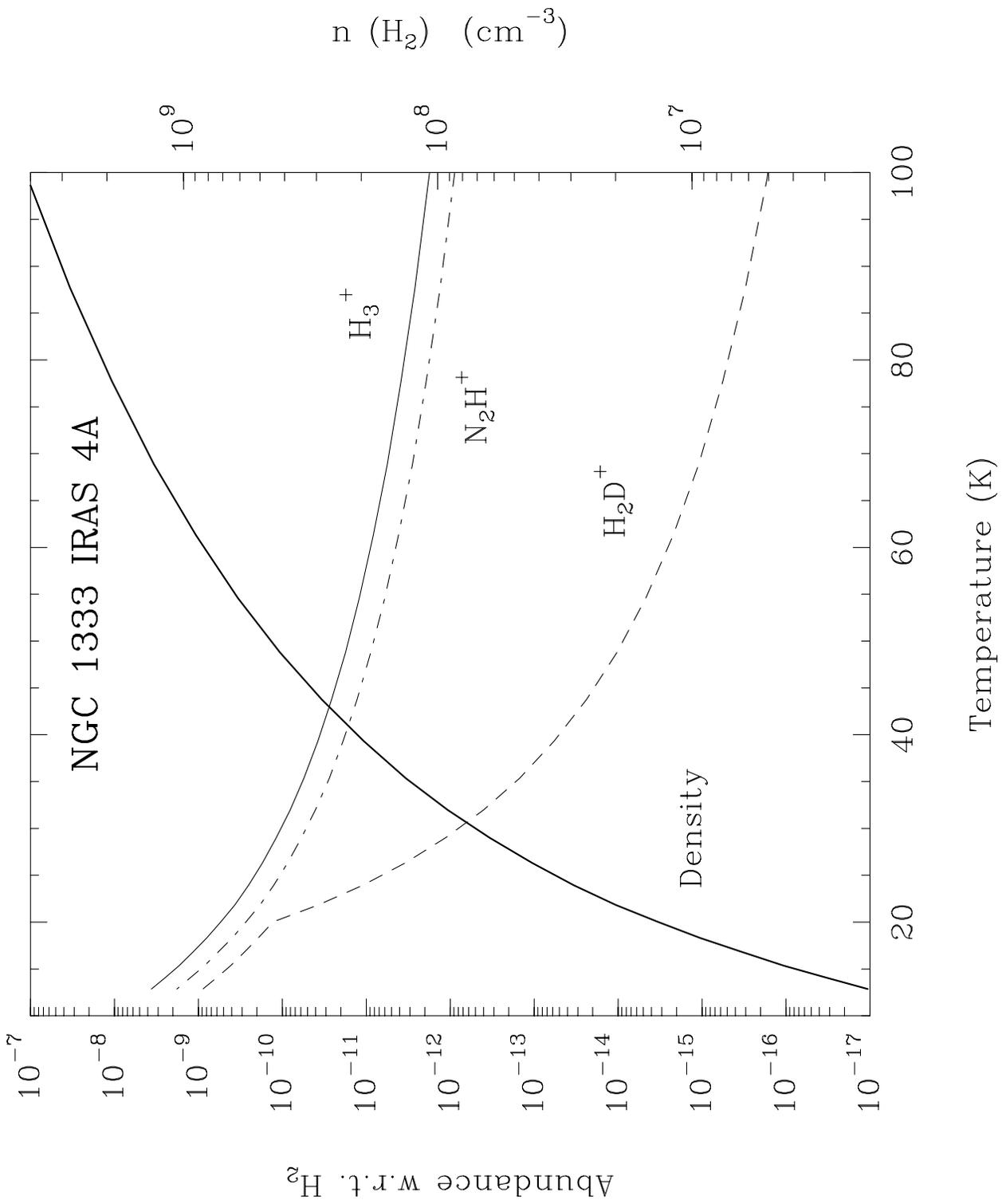
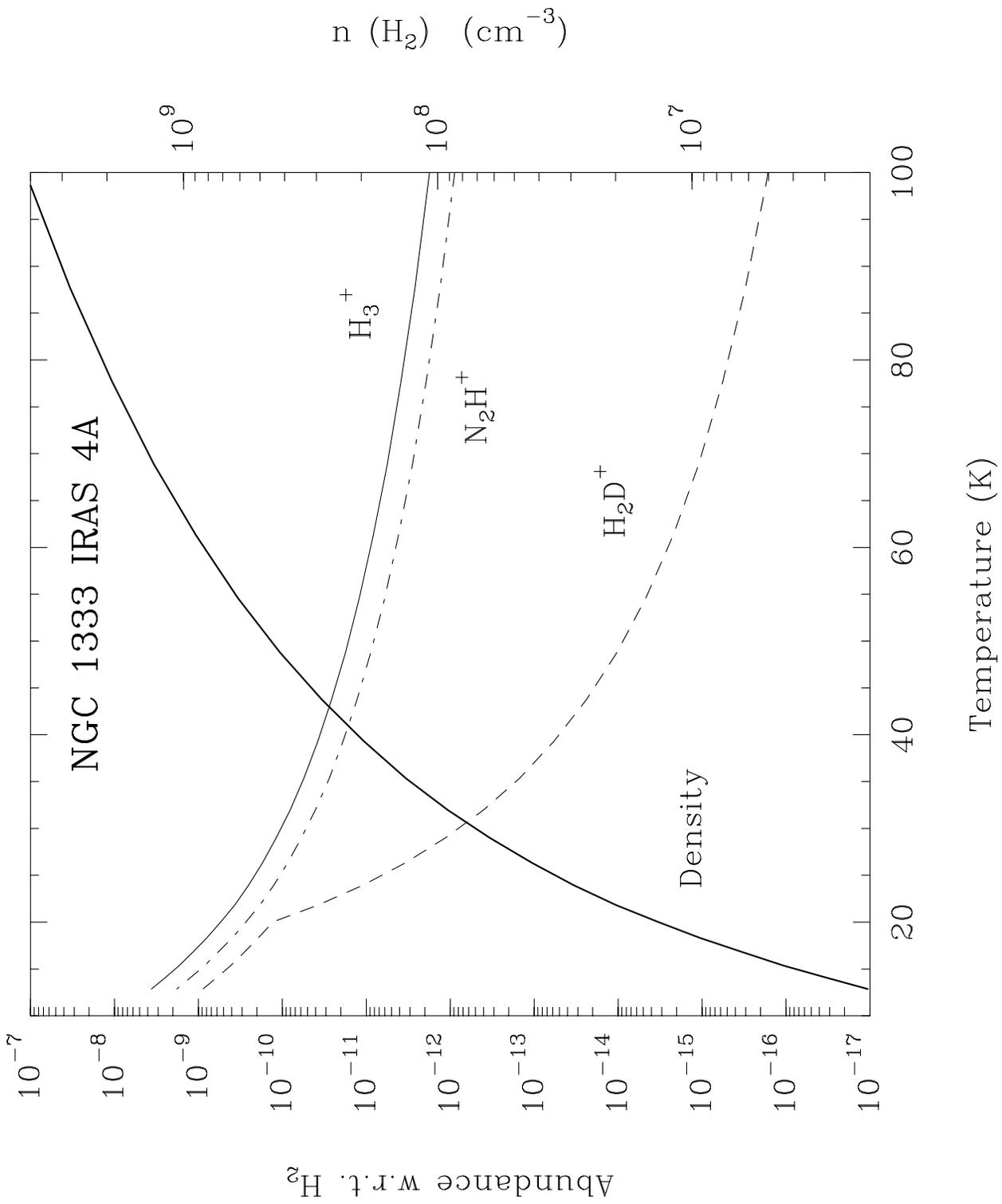
\end{document}